\documentstyle[12pt]{article}

\begin{document}
\setcounter{page}{1} \pagestyle{plain} \vspace{5cm}
\begin{center}
\Large{\bf  Non-Minimal Inflation after WMAP3}\\
\small
\vspace{1cm} {\bf Kourosh Nozari}\quad\ and \quad {\bf S. Davood Sadatian}\\
\vspace{0.5cm} {\it Department of Physics,
Faculty of Basic Sciences,\\
University of Mazandaran,\\
P. O. Box 47416-1467, Babolsar, IRAN\\
knozari@umz.ac.ir,\, d.sadatian@umz.ac.ir}
\end{center}
\vspace{1.5cm}
\begin{abstract}
The Wilkinson Microwave Anisotropy Probe (WMAP) three year results
are used to constraint non-minimal inflation models. Two different
non-minimally coupled scalar field potentials are considered to
calculate corresponding slow-roll parameters of non-minimal
inflation. The results of numerical analysis of parameter space are
compared with WMAP3 data to find appropriate new constraints on the
values of the non-minimal coupling. A detailed comparison of our
results with previous studies reveals the present status of the
non-minimal inflation model after WMAP3.

\end{abstract}
\newpage

\section{Motivation}

Non-minimal coupling (NMC) of scalar field with gravity is necessary
in many situations of physical and cosmological interest. There are
several compelling reasons to include an explicit non-minimal
coupling in the action ( see for example [1,2,3,4] and references
therein ). NMC arises at the quantum level when quantum corrections
to the scalar field theory are considered. It is necessary also for
the renormalizability of the scalar field theory in curved space. In
most theories used to describe inflationary scenarios, it turns out
that a non-vanishing value of the coupling constant is unavoidable
[2]. In general relativity, and in all other metric theories of
gravity ( in which the scalar field is not part of the gravitational
sector ), the coupling constant necessarily assumes a non-vanishing
value[5]. The study of the asymptotically free theories in an
external gravitational field with a Gauss-Bonnet term shows a scale
dependent coupling parameter. For instance, asymptotically free
grand unified theories have a non-minimal coupling depending on a
renormalization group parameter that converges to the value of
$\frac{1}{6}$ or to any other initial conditions depending on the
gauge group and on the matter content of the theory[6]. In view of
the above results and several other motivations( see for example
[7,8] and references therein), it is then natural to incorporate an
explicit NMC between scalar field and Ricci scalar in the
inflationary paradigm. Generally, with non-minimally coupled scalar
field it is harder to achieve accelerated expansion of the
universe[2,7]. Part of this difficulty is related to the more
sophisticated machinery of fine tuning. Nevertheless, over the last
decade several non-minimal inflation scenario have been proposed to
find reliable framework for issues such as graceful exit of
inflationary phase and the observational constraints on the values
that non-minimal coupling can attain in order to have successful
inflationary scenario are discussed [8-16].

The recent astronomical observations, specially high precision data
of WMAP3 [17] have opened new doors in the field of observational
cosmology. As a result, these data have significant impact on the
inflation paradigm. In this regard, these high precision data will
give more accurate bounds on the values of non-minimal coupling in a
typical non-minimal inflation model. The purpose of this letter is
to study impact of WMAP3 and non-minimal inflation. Considering some
well-known inflationary potentials, we explore new observational
constraints on the values of non-minimal coupling to have successful
non-minimal inflation. By definition of an effective scalar field
potential, we show that there is a region in parameter space that
inflation is driven by the non-minimal coupling term. A detailed
study of spectral index and its running shows the spontaneous exit
of inflationary phase ( without any mechanism ) in a suitable region
of the parameter space. We also compare our results with the results
of previous studies. This comparison reveals the present status of
non-minimal inflation paradigm after WMAP3.

\section{Non-Minimal Inflation}

To construct a specific non-minimal inflation scenario, we start
with the following action in Jordan frame
\begin{eqnarray}
S=\int d^4x \sqrt{-g} \left[ \frac{1}{2\kappa^2}R
-\frac{1}{2}g^{\mu\nu}\partial_{\mu}\phi \partial_{\nu}\phi
+\frac{1}{2} \xi R \phi^2+V(\phi)\right]
\label{1}
\end{eqnarray}
where $\frac{\kappa^{2}}{8\pi} \equiv G =m_{\rm pl}^{-2} $ is
Newton's gravitational constant, and $\xi$ is a coupling constant.
The metric signature convention is chosen to be $(+\,-\,-\,-)$ with
spatially flat Robertson-Walker metric as follows
\begin{eqnarray}
ds^2=dt^2-a^{2}(t)\delta_{ij}dx^idx^j.
 \label{2}
\end{eqnarray}
To obtain the fundamental background equations in Einstein frame, we
perform the following conformal transformation to the Einstein frame
\begin{eqnarray}
\hat{g}_{\mu\nu}=\Omega g_{\mu\nu}, ~~~~~~
\Omega=1+\kappa^2\xi\phi^2. \label{3}
\end{eqnarray}
Here we use a hat on a variable defined in the Einstein frame. The
conformal transformation gives
\begin{eqnarray}
S=\int d^4x \sqrt{-\hat{g}} \left[ \frac{1}{2\kappa^2}\hat{R}
-\frac{1}{2}F^2(\phi)\hat{g}^{\mu\nu}\partial_{\mu}\phi\partial_{\nu}\phi
+\hat{V}(\phi) \right], \label{4}
\end{eqnarray}
where by definition
\begin{eqnarray}
F^2(\phi)\equiv\frac{1+\kappa^2\xi\phi^2(1+6\xi)}{(1+\kappa^2\xi\phi^2)^2}
\label{5}
\end{eqnarray}
and
\begin{eqnarray}
\hat{V}(\phi)\equiv\frac{V(\phi)}{(1+\kappa^2\xi\phi^2)^2}.
\label{6}
\end{eqnarray}
Therefore, one may redefine the scalar field as follows
\begin{eqnarray}
\frac{d\hat{\phi}}{d\phi}=F(\phi)=\frac{\sqrt{1+\kappa^2\xi\phi^2(1+6\xi)}}{1+\kappa^2\xi\phi^2}.
\label{7}
\end{eqnarray}
When we investigate the dynamics of universe in the Einstein frame,
we should transform our coordinates system to make the metric in the
Robertson-Walker form
\begin{eqnarray}
\hat{a}=\sqrt{\Omega}a,~~d\hat{t}=\sqrt{\Omega}dt \label{8},
\end{eqnarray}
and we obtain
\begin{eqnarray}
d\hat{s}^2=d\hat{t}^2-\hat{a}^2(\hat{t})\delta_{ij}dx^idx^j.
\label{9}
\end{eqnarray}
Note that the physical quantities in Einstein frame should be
defined in this coordinate system. Now the Einstein equations can be
written as follows
\begin{eqnarray}
\hat{H}^2=\frac{\kappa^2}{3}\left[
\bigg(\frac{d\hat{\phi}}{d\hat{t}}\bigg)^2+\hat{V}(\hat{\phi})
\right] \label{10},
\end{eqnarray}
\begin{eqnarray}
\frac{d^2\hat{\phi}}{d\hat{t}^2}+3\hat{H}\frac{d\hat{\phi}}{d\hat{t}}+
\frac{d\hat{V}}{d\hat{\phi}}=0. \label{11}
\end{eqnarray}
Using the slow-roll approximations \, $\dot{\hat{\phi}}^2\ll
\hat{V}$ \, and \, $\ddot{\hat{\phi}}\ll 3\hat{H}\dot{\hat{\phi}}
$\, , \, we obtain from $(\ref{10})$ and $(\ref{11})$ the following
dynamical equations
\begin{eqnarray}
\hat{H}^2=\frac{\kappa^2}{3}\hat{V}(\hat{\phi})
\label{12},
\end{eqnarray}
\begin{eqnarray}
3\hat{H}\frac{d\hat{\phi}}{d\hat{t}}+\frac{d\hat{V}}{d\hat{\phi}}=0.
\label{13}
\end{eqnarray}

In which follows we will concentrate on two well-known inflationary
potentials: models with inflation potential of the form \,
$V(\phi)=\lambda\phi^n$\, and models with an exponential potential
\,
$V(\phi)=V_{0}\exp\bigg(-\sqrt{\frac{16\pi}{pm^2_{pl}}}\phi\bigg)$.
With these potentials, we obtain
\begin{eqnarray}
\hat{H}^2\approx \frac{\kappa^2}{3}
\frac{\lambda\phi^n}{(1+\kappa^2\xi\phi^2)^2} \label{14},
\end{eqnarray}
\begin{eqnarray}
\hat{H}^2\approx
\frac{\kappa^2}{3}\frac{V_{0}\exp\bigg(-\sqrt{\frac{16\pi}{pm^2_{pl}}}\phi\bigg)}
{(1+\kappa^2\xi\phi^2)^2}. \label{15}
\end{eqnarray}
respectively. Figure 1 shows the behavior of effective potentials as
defined by relation (6). For $V(\phi)=\lambda\phi^n$\, it is
possible to achieve inflationary phase with appropriate $\xi$. For
$V(\phi)=V_{0}\exp\bigg(-\sqrt{\frac{16\pi}{pm^2_{pl}}}\phi\bigg)$
however, when $\xi$ is negative, since the effective potential is
flat in region of $\phi >0$ compared with the $\xi=0$ case, we can
expect assisted inflation to occur in this region. When $\xi$ is
positive, inflation can be realized in the region of $\phi <0$. For
a detailed study of these issues for power-law inflation see [18].\\

Note that we have used a well-know conformal transformation to
express the action in the Einstein frame. In this frame, the
gravitational sector is expressed in terms of a re-scaled scalar
field which is minimally coupled and evolves in a re-scaled
potential, thereby simplifying the formalism. However, one has to
keep in mind that the matter sector is strongly affected by such a
conformal transformation since all of the matter fields are now
non-minimally coupled to the re-scaled metric: in particular, stress
tensor conservation in the matter sector is no longer ensured
[19,20]. On the other hand, as Makino and Sasaki [21] and Fakir,
Habib and Unruh [22] have proved, the amplitude of scalar
perturbation in the Jordan frame exactly coincides with that in the
Einstein frame. This proof (with a complete details in [23]) allows
us to calculate the scalar power spectrum in the Jordan and Einstein
frame. As a result, the scalar power spectrum has not dependence on
the choice of frames, that is, it is conformally invariant. So, our
results can be compared to observations directly without any
ambiguities[23]. This is an important point since one has to check
validity of non-gravity experiments in Einstein frame. For instance,
validity of electro-magnetic related experiments such as CMB
experiment should be checked in Einstein frame. As Komatsu and
Futamase have shown, the scalar power spectrum is independent on the
choice of frames.

Now we study the effect of non-minimal coupling of scalar field and
curvature on the spectral index of density perturbations.

\begin{figure}
\begin{center}\includegraphics{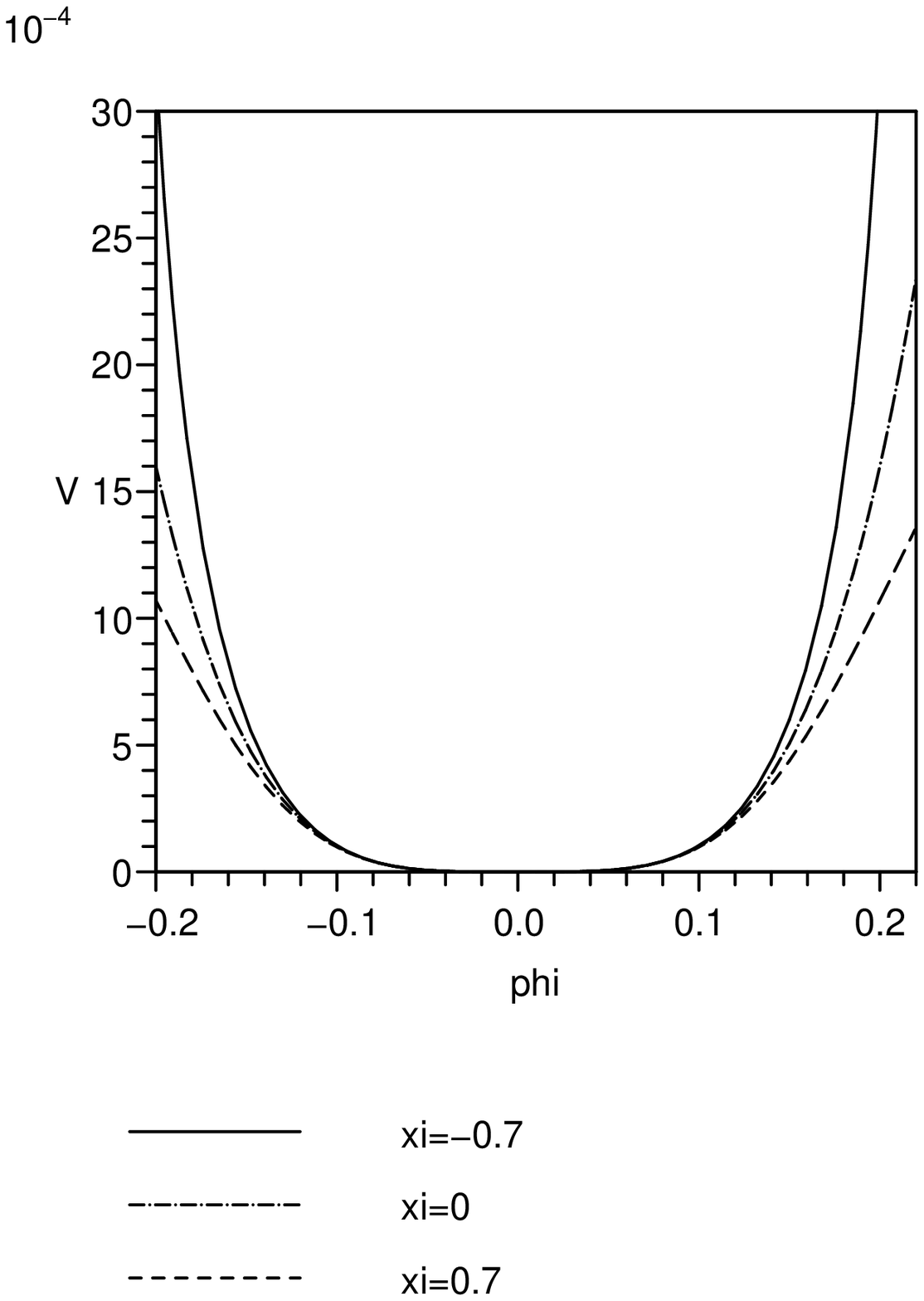} \vspace{6.5cm}\includegraphics{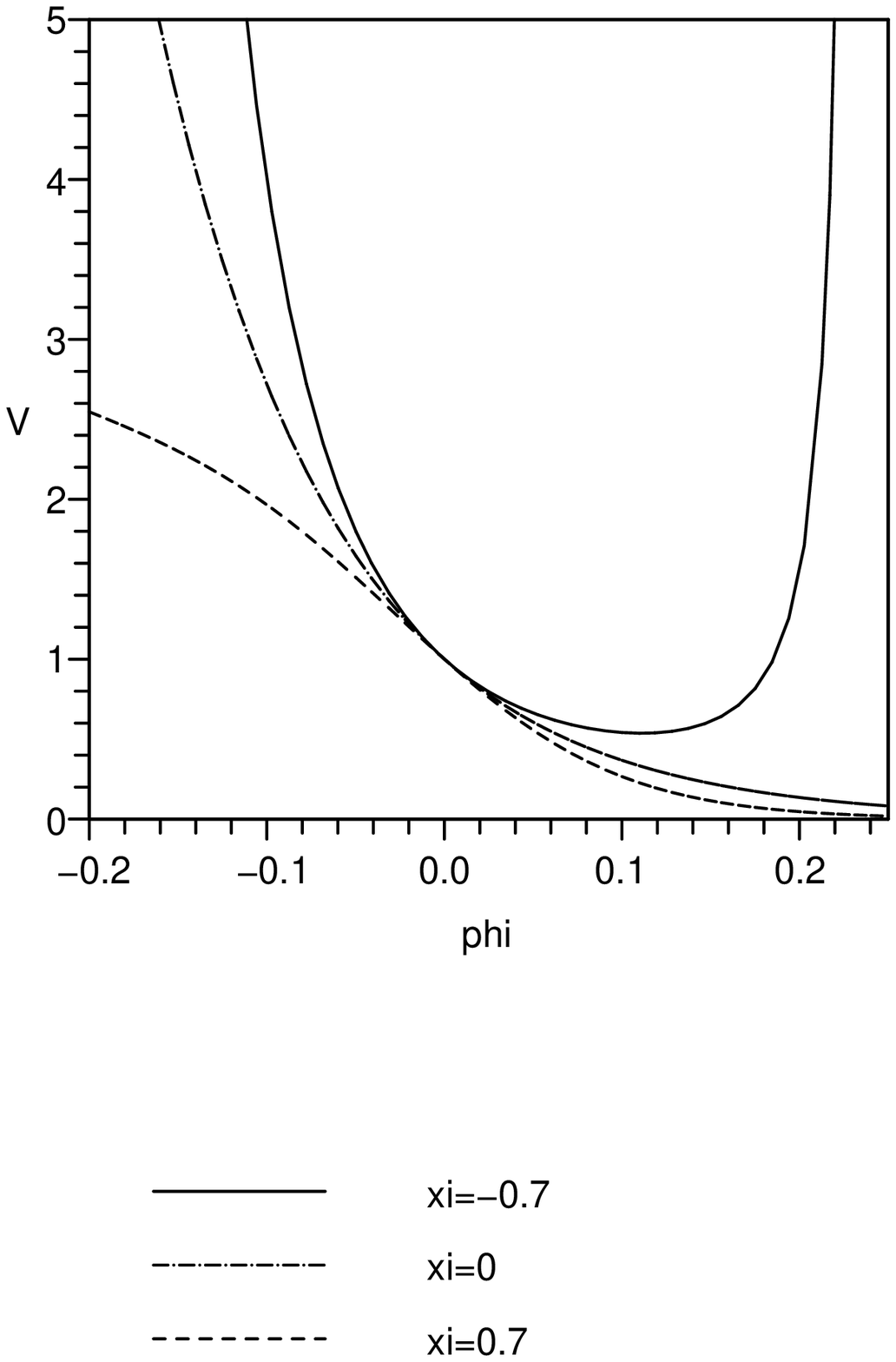}
\end{center}
\caption{\small {Variation of~ $\hat{V}$ relative to $\phi$ for
different values of $\xi$ for potential $\lambda\phi^n$ with $n=4$
(left) and exponential potential (right). }} \label{fig:1}
\end{figure}

\subsection{Slow-Roll Parameters}

In this section we calculate the spectral index of density
perturbations in our non-minimal framework. As Komatsu and Futamase
have shown, in the first order approximation in slow-roll
parameters, $n_s$ is invariant under the conformal transformation
[23]. We use the following definitions of slow-roll parameters
versus $ \hat{\phi} $ as defined by Liddle $et~ al$[24] (see also
[25-27]):
\begin{eqnarray}
\epsilon\equiv\frac{1}{2\kappa^2}\Bigg(\frac{\hat{V}'(\hat{\phi})}{\hat{V}
(\hat{\phi})}\Bigg)^2~~~~~,~~~~~\eta\equiv\frac{1}{\kappa^2}\Bigg(\frac{\hat{V}''(\hat{\phi})}
{\hat{V}(\hat{\phi})}\Bigg), \label{16}
\end{eqnarray}
and
\begin{eqnarray}
\zeta\equiv\frac{1}{\kappa^2}\Bigg(\frac{\hat{V}'(\hat{\phi})\hat{V}'''(\hat{\phi})}
{\hat{V}^2(\hat{\phi})}\Bigg)^{\frac{1}{2}}. \label{17}
\end{eqnarray}
where a prime denotes $\frac{d}{d\hat{\phi}}$. These parameters can
be related directly to observable $~
n_s-1=-6\epsilon+2\eta$\,\,[24-27]. Of course, to the second order
of slow-roll parameters, the spectral index will attain different
form[25]. Now we use equation $(\ref{16})$, $(\ref{6})$ and
$(\ref{7})$  to determine $~n_s~$. Since,
$\hat{V}'=\frac{d\hat{V}}{d\hat{\phi}}=\frac{d\phi}{d\hat{\phi}}\frac{d\hat{V}}{d\phi}
$, we find

\begin{eqnarray}
\epsilon=\frac{1}{2\kappa^2\phi^2}\frac{[n+\kappa^2\xi\phi^2(n-4)]^2}
{[1+\kappa^2\xi\phi^2(1+6\xi)]},
\label{18}
\end{eqnarray}
$$\eta=\frac{1}{\kappa^2}\frac{1}{(1+\kappa^2\xi\phi^2)^2}
\bigg[(1+\kappa^2\xi\phi^2)(n(n+1)\phi^{-2}+n(n+3)\kappa^4\xi^2\phi^2+2n(n+1)
\kappa^2\xi-4(n+1)\kappa^2\xi$$
$$-4(n+3)\kappa^4\xi^2\phi^2-
8\kappa^2\xi(n+n\kappa^4\xi^2\phi^4+2n\kappa^2\xi\phi^2-
4\kappa^2\xi\phi-4\phi^3\kappa^4\xi^2)
(\frac{1}{(1+\kappa^2\xi\phi^2(1+6\xi))^{\frac{1}{2}}})\bigg]$$
$$+\frac{1}{\kappa^2}\times\frac{2\kappa^2\xi\phi
(1+\kappa^2\xi\phi^2(1+6\xi))^{\frac{1}{2}}-(1+2\kappa^2\xi\phi^2)
\kappa^2\xi\phi(1+6\xi)(1+\kappa^2\xi\phi^2(1+6\xi))^{\frac{-1}{2}}}
{(1+\kappa^2\xi\phi^2(1+6\xi))}$$
\begin{eqnarray}
\times\frac{(n\phi^{-1}(1+\kappa^2\xi\phi^2)^2-4\kappa^2\xi\phi(1+\kappa^2\xi\phi^2))}
{(1+\kappa^2\xi\phi^2)^2}.
\label{19}
\end{eqnarray}
for\,$~V(\phi)=\lambda\phi^n$. For potential
$V(\phi)=V_0\exp(-\sqrt{\frac{16\pi}{pm^2_{pl}}}\phi)$ we derive
\begin{eqnarray}
\epsilon=\frac{1}{2\kappa^2}\left(\frac{-4\xi\kappa^2\phi-
\sqrt{\frac{16\pi}{pm^2_{pl}}}(1+\xi\kappa^2\phi^2)}
{\sqrt{1+(1+6\xi)\xi\kappa^2\phi^2}}\right)^2,
\label{20}
\end{eqnarray}
$$\eta=\frac{1}{2\kappa^2}\Bigg(\frac{(-4\xi\kappa^2+4\xi\kappa^2\phi
\sqrt{\frac{16\pi}{pm^2_{pl}}})+(-2\xi\kappa^2\phi\sqrt{\frac{16\pi}{pm^2_{pl}}}+
\frac{16\pi}{pm^2_{pl}}(1+\xi\kappa^2\phi^2))}{(1+\xi\kappa^2\phi^2)^2
(1+\xi\kappa^2\phi^2(1+6\xi))}-$$
$$\bigg[\frac{(-4\xi\kappa^2\phi-\sqrt{\frac{16\pi}{pm^2_{pl}}}(1+\xi\kappa^2\phi^2))
(4\xi\kappa^2\phi(1+\xi\kappa^2\phi^2)(1+\xi\kappa^2\phi^2(1+6\xi))}
{(1+\xi\kappa^2\phi^2)^2(1+\xi\kappa^2\phi^2(1+6\xi))}$$
\begin{eqnarray}
+\frac{(1+6\xi)\xi\kappa^2\phi(1+(1+6\xi)\xi\kappa^2\phi^2)^{\frac{-1}{2}}
(1+\xi\kappa^2\phi^2)^2}{(1+\xi\kappa^2\phi^2)^2(1+\xi\kappa^2\phi^2(1+6\xi))}\bigg]\Bigg).
 \label{21}
\end{eqnarray}
The inflationary phase terminates if the condition $\epsilon=1$ be
satisfied[24]. For exponential inflationary potential as described
above, slow-roll equations cannot be solved in algebraic way, so we
try to solve them numerically. In the case of the large field models
$\lambda\phi^n$, these equations can be solved analytically, however
the result of numerical calculation can be interpreted more easily.
Figures 2 shows the results of these calculations for potential
$V(\phi)=\lambda\phi^n$\, for both positive and negative scalar
field $\phi$. We see that with positive $\xi$, both positive and
negative $\phi$ can lead to spontaneous exit of inflationary phase
without any additional mechanism. For negative $\xi$ however, as
figure 3 shows, the situation is very different. With
$V(\phi)=\lambda\phi^4$ and negative $\phi$ as figure 3 (left)
shows, inflationary phase can exit only for $\xi\leq -0.1666$. For
$\xi\leq -0.1666$ inflationary phase can exit spontaneously without
any mechanism. For all  $\xi > -0.1666$ we find a negative
$\epsilon$ which is evidently impossible since $\epsilon$ is a
positive definite quantity. On the other hand, with positive $\phi$
and negative $\xi$ as figure 3 (right) shows, inflationary phase can
exit spontaneously only for $\xi\leq -\frac{1}{6}$. For
$V(\phi)=V_{0}\exp\bigg(-\sqrt{\frac{16\pi}{pm^2_{pl}}}\phi\bigg)$,
figures 4 and 5 show that inflationary phase can exit spontaneously
with all possible values of $\xi$ regardless of $\phi$ sign. So,
non-minimal coupling provides a successful machinery for spontaneous
exit of inflationary phase.

\begin{figure}[htp]
\begin{center}\includegraphics{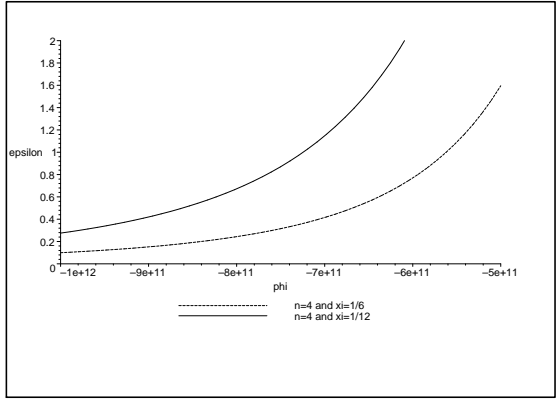} \vspace{5cm}\includegraphics{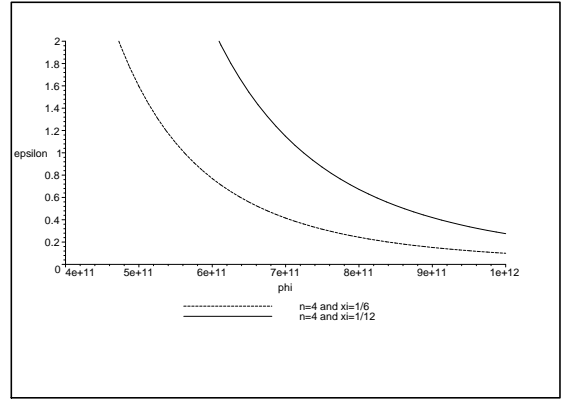}
\end{center}
 \caption{\small {Variation of $\epsilon$
for different values of (positive) non-minimal coupling for
potential $\lambda\phi^n$ with $n=4$. In both cases, it is possible
to exit inflationary phase without any additional mechanism.}}
\label{fig:2}
\end{figure}

\begin{figure}
\begin{center}\includegraphics{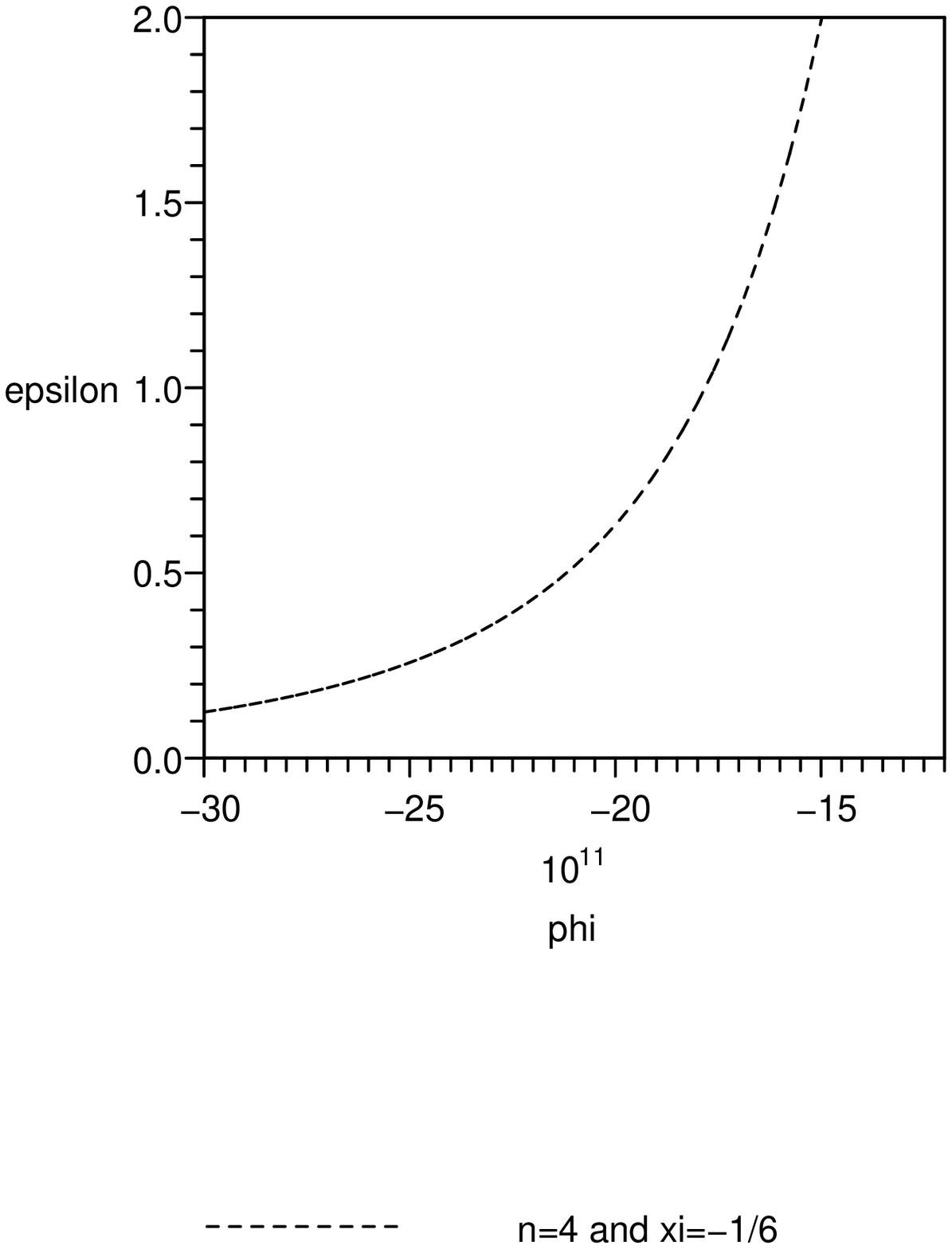} \vspace{10cm}\includegraphics{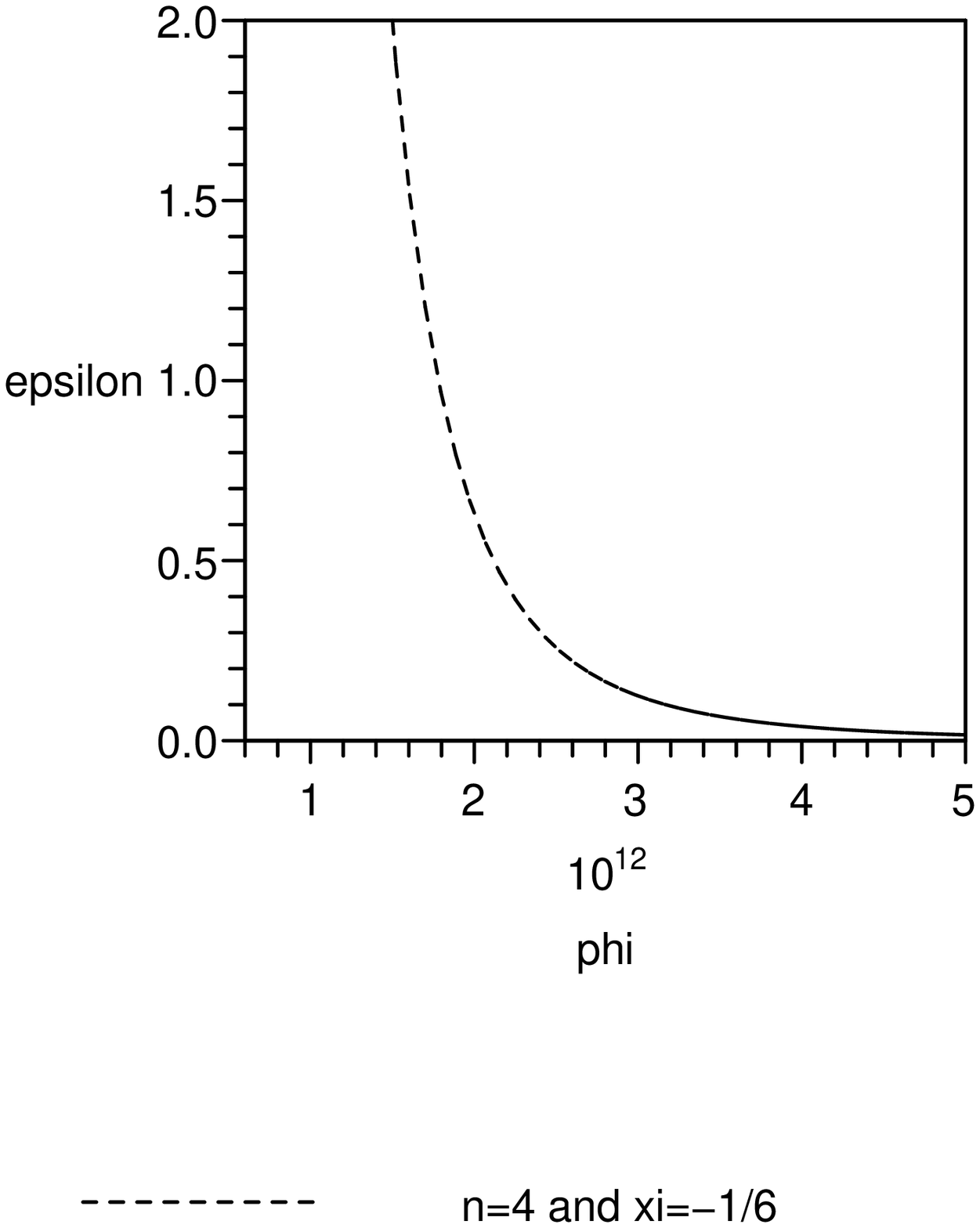}
\end{center}
\caption{\small {Variation of $\epsilon$ for different values of
(negative) non-minimal coupling for potential $\lambda\phi^n$ with
$n=4$. In both cases, it is possible to exit inflationary phase
spontaneously only with suitable range of non-minimal coupling. }}
\label{fig:1}
\end{figure}

\begin{figure}[htp]
\begin{center}\includegraphics{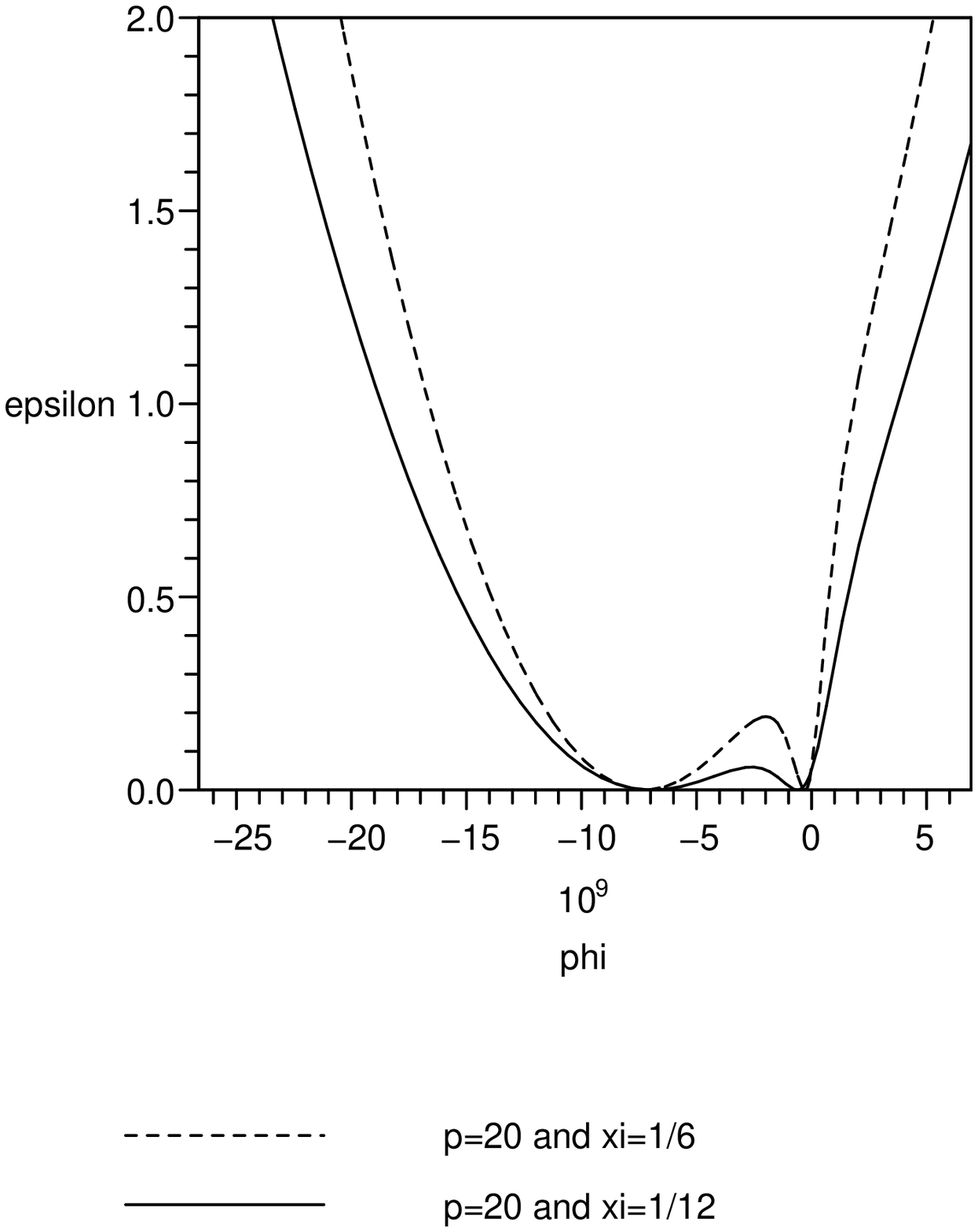} \vspace{6cm}
\end{center}
 \caption{\small {Variation of $\epsilon$ for different values of
 (positive) non-minimal coupling for
potential
$V(\phi)=V_{0}\exp\bigg(-\sqrt{\frac{16\pi}{pm^2_{pl}}}\phi\bigg)$.
In both cases, it is possible to exit inflationary phase without any
additional mechanism.}} \label{fig:2}
\end{figure}

\begin{figure}[htp]
\begin{center}\includegraphics{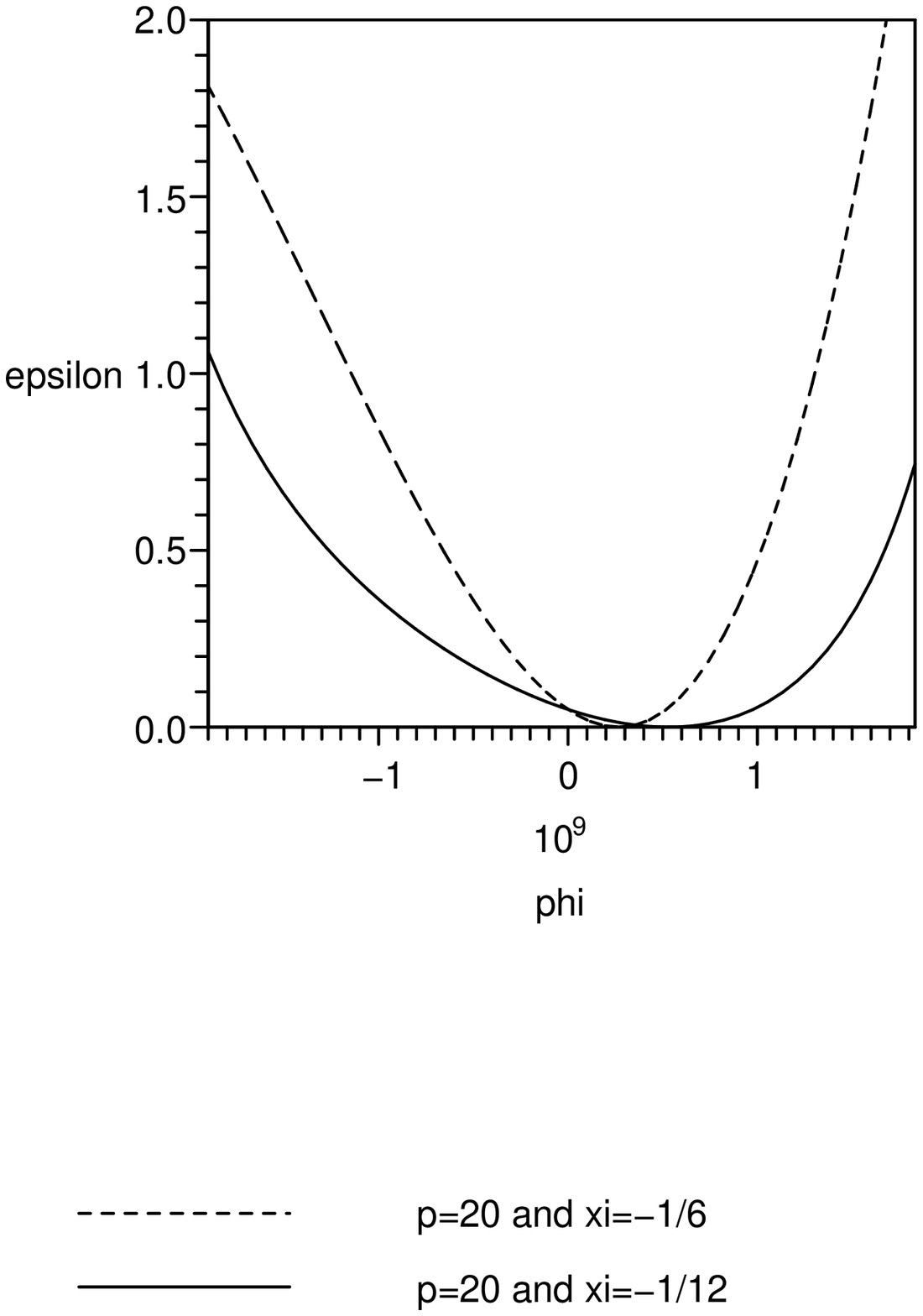} \vspace{9cm}
\end{center}
 \caption{\small {Variation of $\epsilon$ for different values of (negative) non-minimal coupling for
potential
$V(\phi)=V_{0}\exp\bigg(-\sqrt{\frac{16\pi}{pm^2_{pl}}}\phi\bigg)$.
In both cases, it is possible to exit inflationary phase without any
additional mechanism.}} \label{fig:2}
\end{figure}

The number of e-folds, $N\equiv\ln\frac{a_{e}}{a_{i}}$ can be
written as
\begin{equation}
N=-\int_{\hat{\phi_{i}}}^{\hat{\phi_{e}}}3\hat{H}^{2}
\frac{d\hat{\phi}}{d\hat{V}}d\hat{\phi}=
-\sqrt{\frac{4\pi}{m^2_{pl}}}\,\,\bigg|\int_{\hat{\phi_{i}}}^{\hat{\phi_{e}}}
\epsilon^{-\frac{1}{2}}d\hat{\phi}\bigg|.
\end{equation}
where $\hat{\phi_{i}}$ denotes the value of scalar field
$\hat{\phi}$ when Universe scale observed today crosses the Hubble
horizon during inflation, while $\hat{\phi_{e}}$ is the value of
scalar field when the Universe exits the inflationary phase. The
number of e-folds can be obtained using equations $(\ref{6})$ and
$(\ref{10})$. Now we set $n=4$ and for simplicity we choose
$\lambda=\frac{\lambda'}{4}$ in $V(\phi)=\lambda\phi^n$. Writing\,
$\kappa^2\xi\phi^2_{end}=\beta^2(\xi)$\, in relation (18), we can
solve equation $\epsilon=1$ for\, $\beta(\xi)$ to find
\begin{eqnarray}
\beta=\sqrt{\frac{1}{2(1+6\xi)}(\sqrt{192\xi^2+32\xi+1}-1)}.
\label{23}
\end{eqnarray}
If we similarly write $\kappa^2\xi\phi^2_{hc}=m^2(\xi)$ where
$\phi_{hc}$ is the value of $\phi$ when the scales of interest
crossed outside of horizon during inflation, then we can rewrite
relations $(\ref{18})$ and $(\ref{19})$ using the fact that scales
of interest to us crossed outside of horizon is approximately $70$
e-folds before the end of inflation, that is,
$e^N\equiv\frac{\hat{a}(\hat{t}_{end})}{\hat{a}(\hat{t}_{hc})}\sim
e^{70}$. The first-order result,\, $n_s=1-6\epsilon+2\eta$, can be
written approximately, in the limit of  $m\gg 1$, as follows
\begin{eqnarray}
n_s\simeq 1-\frac{32\xi}{16\times70\xi -1}. \label{24}
\end{eqnarray}
The running of the spectral index is therefore
\begin{eqnarray}
\alpha_s=\frac{dn_{s}}{d\ln
k}=16\epsilon\eta-24\epsilon^2-2\zeta\simeq-2\zeta\approx -10^{-2}
\label{25}
\end{eqnarray}
where $k=\hat{a}\hat{H}$. This quantity could be larger and match
observation if we adopt a larger value of the coupling constant
$\lambda$ (for example $\alpha_s=-0.021$ if $\lambda=0.30)$. Figure
6 shows the result of this calculation. As we see, both of these
figures show that $n_{s}\leq 1$ and therefore our model favors a red
and nearly scale invariant spectrum.

\begin{figure}[htp]
\begin{center}\includegraphics{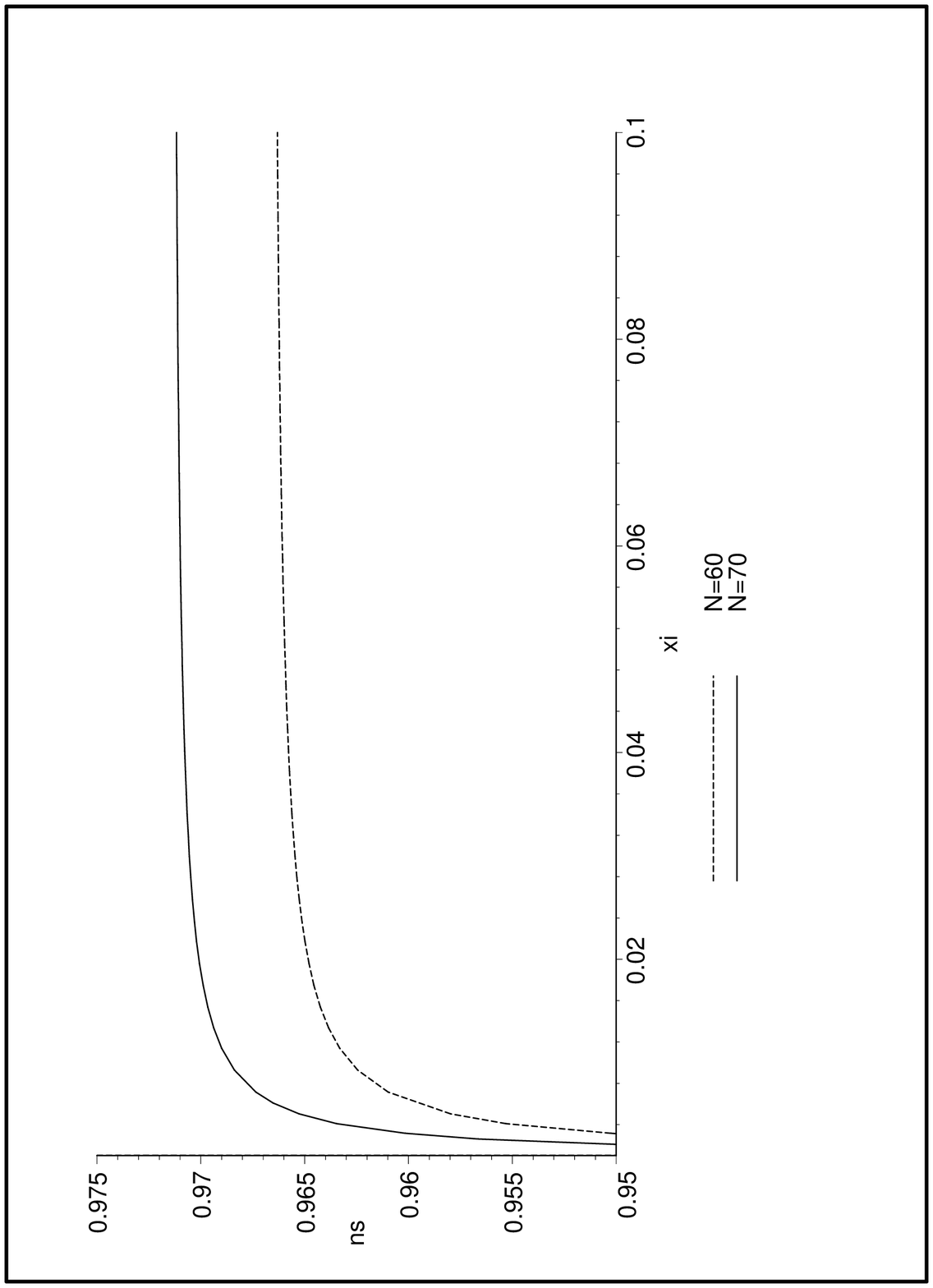}
\vspace{5cm}\includegraphics{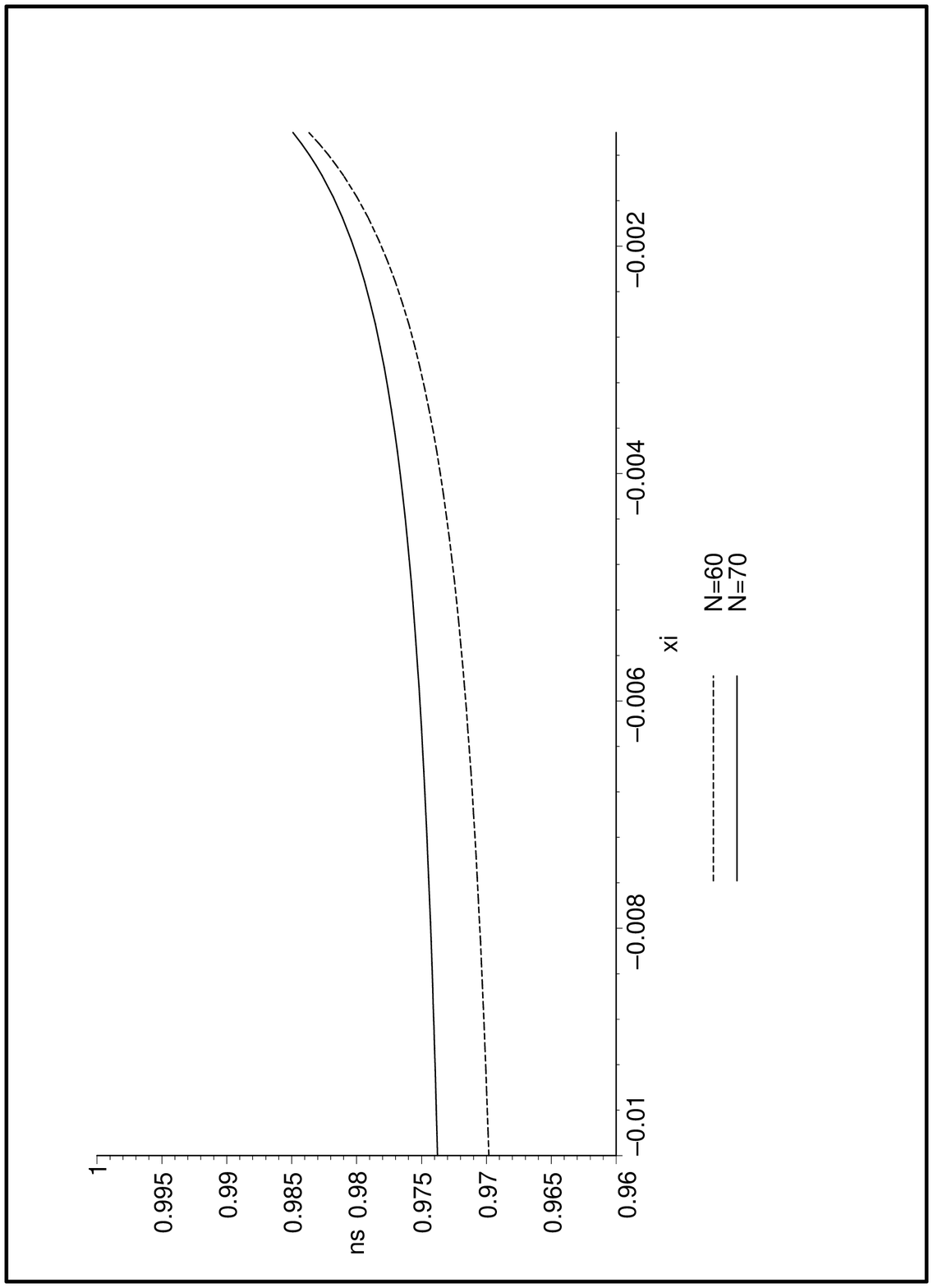}
\end{center}
 \caption{\small {Variation of spectral index, $n_s$
for different values of non-minimal coupling for potential
$\lambda\phi^n$. Evidently, $n_s\leq 1$ and therefore our
non-minimal model favors a red power spectrum. }} \label{fig:1}
\end{figure}
We can similarly calculate $n_s$ and the running of the spectral
index for exponential potential. The results are as follows
\begin{eqnarray}
n_s=1-\frac{3\bigg(-4a\sqrt{\xi}-(1+a)\sqrt{\frac{2}{p}}\bigg)^2}{1+(1+6\xi)a},
\label{26}
\end{eqnarray}
where $p\gg 2$ and $a\sim -\frac{m_{pl}}{\sqrt{8\pi\xi}}$.\, The
running of the spectral index is
\begin{eqnarray}
\alpha_s=\frac{dn_{s}}{d\ln
k}=16\epsilon\eta-24\epsilon^2-2\zeta\simeq-2\zeta\approx
-10^{-1},\label{27}
\end{eqnarray}
for example $\alpha_s=-0.102$ if $p=14$. In this case we cannot
consider $\xi<0$, since with $\xi<0$ relation $(\ref{26})$ for
spectral index becomes a complex quantity. Figure (7) shows the
variation of spectral index for exponential potential. We see that
in this case we have always $n_{s}\leq 1$. Therefore, our model for
exponential potential gives also a nearly scale invariant spectrum.

\begin{figure}[htp]
\begin{center}\includegraphics{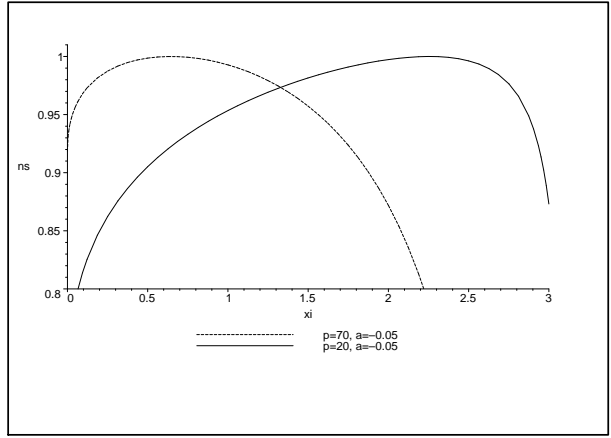} \vspace{5cm}
\end{center}
 \caption{\small {Variation of $n_s$
for different values of non-minimal coupling for potential
$V(\phi)=V_{0}\exp\bigg(-\sqrt{\frac{16\pi}{pm^2_{pl}}}\phi\bigg)$.
}} \label{fig:1}
\end{figure}

\section{Numerics of Parameter Space and Observational Constraints}
To study cosmological implications of this non-minimal inflation
model, we perform some numerical analysis of parameter space. The
result of WMAP3 for $\Lambda$CDM gives
$n_{s}=0.951^{+0.015}_{-0.019}$ for index of the power spectrum[17].
Combining WMAP3 with SDSS (Sloan Digital Sky Survey), gives
$n_{s}=0.948^{+0.015}_{-0.018}$ at the level of one standard
deviation[12]. These results show that a red power spectrum is
favored at least at the level of two standard deviations. If there
is running of the spectral index, the constraints on the spectral
index and its running are given by [12]
\begin{equation}
n_{s}=1.21^{+0.13}_{-0.16}
\end{equation}
and
\begin{equation}
\frac{dn_{s}}{d\ln k}=-0.102^{+0.050}_{-0.043}.
\end{equation}
Table 1 summarizes the results of our calculations for spectral
index for two different kind of scalar field potential. This table
contains also the constraints imposed on the non-minimal coupling in
comparison with WMAP3 data. Choosing some limiting values of
$\alpha_{s}$, we obtain a suitable range of non-minimal coupling. In
this table, $a\sim -\frac{m_{pl}}{\sqrt{8\pi\xi}}$ is the value of
the scalar field when the inflationary phase terminates. Therefore,
to have assisted inflation with non-minimal coupling, we should
restrict ourselves to the conditions $\xi\leq -0.1666$ and
$\xi\geq0.01$ for $V(\phi)=\lambda \phi^{n}$ and
$0.271\leq\xi\leq0.791$ for exponential potential. Since the values
that $n_{s}$ can attains in our non-minimal framework are less than
unity, our model shows a red power spectrum in accordance with
WMAP3. On the other hand,
our model shows a nearly scale invariant spectrum.\\
\begin{table}[htp]
\begin{center}
\caption{Constraining Non-minimal inflation with WMAP3} \vspace{0.5
cm}
\begin{tabular}{|c|c|c|c|c|c|c|c|c|c|c|}

\hline
\hline $V(\phi)$&$n_s$&$\alpha_s $&$\xi$ \\
\hline $\frac{\lambda}{4}\phi^4$ &$1-\frac{32\xi}{16\times70\xi
-1}$&
$\sim-10^{-2}$ &$\xi\leq -0.1666$ and $\xi\geq0.01$\\
\hline $V_{0}\exp(-\sqrt{\frac{16\pi}{pm^2_{pl}}}\phi)$&
$n_s=1-\frac{3(-4a\sqrt{\xi}-(1+a)\sqrt{\frac{2}{p}})^2}{1+(1+6\xi)a}$
& $\sim -10^{-1}$&$0.271\leq\xi\leq0.791$ \\
\hline
\end{tabular}
\end{center}
\end{table}
\\
\\
\\
At this stage we compare our results with the previous studies.
Table 2 summarizes the results of previous studies in comparison
with our present results. As we see, our constraints for non-minimal
coupling with exponential potential are consistent with holographic
dark energy framework and also with result of warped DGP brane
inflation. On the other hand, with potential of the form
$V(\phi)=\lambda \phi^{n}$, we find a more precise interval of the
non-minimal coupling for assisted inflation. It is well-known that
multiple scalar fields with exponential potential can lead to an
inflationary solution even if each scalar field alone fails to
provide this situation. Here we see that with non-minimally coupled
scalar field the situation is different and non-minimal coupling
itself can assist the inflationary phase. So, inflation can be
assisted by non-minimal coupling even in the one field case. In this
framework, primordial perturbations are almost scale-independent.
The primordial power spectrum predicted by this non-minimal
inflation is almost scale-independent, that is the spectral index
$n_s$ is very close to unity. Possible deviations from exact scale
independence arise because during inflation the inflaton is not
massless and the Hubble rate is not exactly constant. As we know,
the WMAP3 data favors a red power spectrum at the level of two
standard deviations, which provides a stringent constraint on the
inflation models. While allowing for a running spectral index
slightly improves the fit to the WMAP data. With these
preliminaries, our non-minimal inflation model provides a realistic
framework for spontaneous exit of inflationary phase without any
additional mechanism with suitable constraints imposed on the values
that non-minimal coupling can attains.

In summary, with non-minimal coupling there is a region of parameter
space that inflation is driven by the non-minimal coupling term
alone. Also, non-minimal inflation provides a suitable mechanism for
spontaneous exit of inflationary phase in some special
circumstances. Our non-minimal inflation scenario gives a red and
nearly scale invariant power spectrum. Comparison of non-minimal
inflation model with WMAP3 data gives more accurate constraints on
the values of non-minimal coupling. A detailed comparison between
our results and results of previous studies shows that our
constraints for non-minimal coupling with exponential potential are
consistent with holographic dark energy framework and also with
result of warped DGP brane inflation. On the other hand, with
potential of the form $V(\phi)=\lambda \phi^{n}$, we find a more
precise interval of the non-minimal coupling for assisted inflation.
Since we have calculated the first order contributions in slow-roll
parameters, our results are valid in both Jordan and Einstein frame.
However, higher order corrections evidently lead to different
results in these two frames[23].
\newpage

\begin{table}[htp]
\begin{center}
\caption{Constraints on $\xi$ in different Non-minimal Inflation
Models} \vspace{0.5 cm}
\begin{tabular}{|c|c|c|c|c|c|c|c|c|c|c|}
\hline
\hline $Model$&$Authors$&$Constraint~on~\xi$&$Source$ \\
\hline $ Holographic~dark~energy[10] $&$M. Ito $ &$0.146\leq\xi\leq0.167 $&$HS~as~IR^1$\\
\hline $ Quintessence~Model[11] $&$T. Chiba $ &$-10^{-2}\leq\xi\leq10^{-2}$&$SSE^2$\\
\hline $ Scaler~Field~theory[9]$&$S. Koh~{\it et~ al} $ &$-10^{-3}<\xi<10^{-3} $&$Theoretical$\\
\hline $Quintessence[28] $&$S.M .Carroll $ &$-10^{-4}\leq\xi\leq10^{-4} $&$CRS^3 $\\
\hline $CMB~ Anisotropy[29] $&$T. Futamase{\it~ et~ al} $ &$\xi>10^{-2} $&$CMB$\\
\hline $COBE~Data~Analysis[30] $&$S. Hancock{\it~et~ al} $ &$-10^{-4}\leq\xi\leq-10^{-3}$&$COBE $\\
\hline $Quantum~radiative~Process[31] $&$A. Bilandzic{\it~et~ al} $ &$0.002\ll|\xi|\ll0.041$&$COBE$\\
\hline $Density~perturbations[32] $&$S. Tsujikawa{\it~et~al}$&$-0.007<\xi<-0.0017$&$CMB$\\
\hline $DGP~Brane~Inflation[8] $&$K. Nozari{\it~ et~ al} $ &$\xi\geq 1.2\times10^{-2} $&$WMAP3$\\
\hline $Our~ Model^4$& $K. Nozari {\it~ et~ al}$&$\xi\leq -0.1666~,~\xi\geq0.01$&$WMAP3$ \\
\hline $Our~ Model^5$& $K. Nozari {\it~ et~ al}$&$0.271\leq\xi\leq0.791$&$WMAP3$ \\
\hline
\end{tabular}
\end{center}
$^1$ Hubble Scale as Infra-Red Cutoff\\
$^2$ Solar System Explorer\\
$^3$ Cosmological Radio Sources\\
$^4$ For $V(\phi)=\lambda \phi^{4}$ in tree-level\\
$^5$ For
$V(\phi)=V_{0}\exp\Big(-\sqrt{\frac{16\pi}{pm^2_{pl}}}\phi\Big)$.
\end{table}

\end{document}